\documentclass[a4paper,11pt]{article}
\pdfoutput=1 

\usepackage{jheppub} 

\usepackage[multidot]{grffile}

\usepackage[T1]{fontenc} 

\usepackage{tikz}

\newcommand{\beq}{\begin{equation}}
\newcommand{\eeq}{\end{equation}}

\title{\boldmath Restoring rotational invariance for lattice QCD propagators.}

\author[a]{F. de Soto}
\affiliation[a]{Departamento de Sistemas F\'\i sicos, Qu\'\i micos y Naturales, Universidad Pablo de Olavide,\\ Ctra. Utrera km 1, 41013, Spain.}
\emailAdd{fcsotbor@upo.es}

\abstract{
	This note presents a method to reduce the discretization errors appearing when solving a Quantum Field Theory in a hypercubic lattice in both position and momentum-space. 
	The method exploits the artifacts that break rotational symmetry to recover rotationally invariant results for two-point Green functions. We show that a combination of the results obtained in position and momentum space can be useful to signal the presence of rotationally invariant artifacts making use of their approximate Fourier transforms in the continuum. 
	The method will be introduced using a Klein-Gordon propagator, and a direct application to gluon propagator in quenched lattice QCD will be given.}

\begin{document} 
\maketitle
\flushbottom

\section{Introduction}
\label{sec:intro}

The presence of discretization errors in lattice field theories has been largely studied since the beginning of lattice QCD. Indeed, when space-time is discretized in an homogeneous lattice of spacing $a$, any observable will --in general-- have errors of order $\mathcal{O}(a^n)$, with $n=1,2,\cdots$ depending on the lattice  action and operator chosen. In this line, Symanzik designed a systematic method to reduce discretization errors where improved operators and lattice actions~\cite{Symanzik:1983dc,Symanzik:1983gh,Luscher:1996sc,Luscher:1996ug} are used to eliminate the dominant counterparts, thus increasing the order, $n$, of the discretization errors present.
Discretization errors disappear once the continuum limit is taken, i.e., by computing the desired observable  at different lattice spacings and extrapolating to the $a\to 0$ limit.

A particularly vexing effect of discretizing space-time in a square lattice is rotational invariance breaking: while the continuum (Euclidean) space-time in $d$-dimensions is invariant under the rotations group~\footnote{A reminiscent of the Lorentz invariance in Minkowsky space.} $SO(d)$, in the lattice this symmetry is broken to the discrete isometry group $H(d)$. As a consequence, lattice propagators in momentum space instead of being smooth functions of $k^2$ acquire  a spurious (in the sense that it is not expected in the continuum) dependence on higher order invariants of the lattice isometry group.
This type of discretization error is responsible for the known half-fishbone structure observed in the gluon or quark propagators when one considers all the lattice momenta~\cite{Becirevic:1999hj,Becirevic:1999uc,Boucaud:2003dx,Bonnet:2001uh} and, over the years,  different strategies have been employed to mitigate the impact of rotational invariance breaking over the lattice results. Among them, choosing only some particular values of the momenta~\cite{Leinweber:1998uu}, using a tree-level correction~\cite{Bonnet:2001uh,Constantinou_2009} or extrapolating to zero the higher order invariants of the lattice group~\cite{deSoto:2007ht,Catumba:2021hcx}. 

The quest for small pion masses and lattice spacings has reached a new era in lattice-QCD in the last decade, with several collaborations achieving pion masses close to the physical value and lattice spacings below $0.1$fm with $N_f=2$, $2+1$ and $2+1+1$ dynamical quark flavours~\cite{FlavourLatticeAveragingGroup:2019iem}. The impact of discretization errors is mitigated by the use of small lattice spacings, but they are still relevant as there is a competition between discretization and statistical errors. As the number of configurations increases, the statistical errors are reduced and discretization ones become more relevant. In any case, the quest for precision measurements reinforces the necessity for a careful treatment of discretization errors.

While the impact of symmetry breaking artifacts in momentum space has been widely studied, for correlation functions in position space one encounters that the effects of breaking the rotational invariance are difficult to deal with. Part of the difficulty is due to the fact that discretization errors more relevant for small distances than for large ones (while in momentum space they are relevant at large momenta). This has been found, for example, in the lattice determination of the quenched potential (See Fig.~4.3 in \cite{Bali:2000gf}), or the renormalization constants for the vector and axial-vector quark bilinears, $Z_V$ and $Z_A$ \cite{Aoki_2004}. In general, symmetry breaking artifacts in position space are not so well understood as in momentum space and are difficult to eliminate.

Discretization artifacts associated to rotational invariance breaking are particularly relevant for observables such as gluon or ghost propagators~\cite{Boucaud:2017ksi,Boucaud:2018xup,Aguilar:2019uob,Aguilar:2021okw}, where statistical errors are well under control. Indeed, the presence of uncontrolled discretization errors in the large-momentum region of the renormalization scale may have an impact on the low-momentum domain. The effect would be similar to the one detected when there is a systematic deviation in the lattice scale setting, as discussed in \cite{Boucaud:2017ksi}.
Although they are also present for the lattice estimate of three-point functions such as the three-gluon vertex, the typically larger statistical errors prevent appreciating the effects of rotational invariance breaking.

In this paper we propose an unified method to deal with hypercubic errors both in position and momentum space based in a generalization of the one presented in \cite{deSoto:2007ht}. Moreover, we will introduce the use of an approximate Fourier transform of the lattice data after appropriate treatment of the hypercubic errors in both position and momentum-space that can be used to detect the presence of  uncorrected O4-invariant discretization or finite-volume errors.

The structure of the paper is as follows: in section \ref{sec:kg} we will use the propagator of a free massive field to discuss the rotational symmetry breaking artifacts in both momentum and position space. Based in this analysis, we will introduce a method to cure them in section \ref{sec:cure} and apply it to the free boson case where the exact solution is known in this section. In section \ref{sec:ft}, we will use a Fourier transform to check the rotational invariance of the extrapolated data. Finally, we will apply this method in section \ref{sec:gluon} to the quenched QCD gluon propagator in position and momentum space for a small lattice.

\section{Analysis of Klein-Gordon propagator}
\label{sec:kg}

In order to get a pedagogical introduction to the problem, we will start by considering the propagator of a free massive boson in momentum space. It has the advantage that it can be straightforwardly obtained in either a discrete or continuum space and both in a finite box or in an infinite space. In this sense, two considerations are in order: i) when working with discrete positions one has 
to make an approximation on how derivatives are written in the discrete space, and ii) dealing with finite volumes implies making some assumption on the boundary conditions. For the cases illustrated here, we considered the simplest cases of forward derivative
\begin{eqnarray}\label{eq:forward}
\frac{\partial \psi(x)}{\partial x} \to \frac{\psi(x+a)-\psi(x)}{a}\ ,
\end{eqnarray}
and periodic boundary conditions 
\begin{eqnarray}
\psi(x+L)=\psi(x)\ ,
\end{eqnarray}
respectively. We will also focus in the four-dimensional case and assume through the manuscript that both the lattice spacing, $a$, and the number of lattice sites, $N$, are equal for all the four space-time directions. While the former is a  common choice in lattice QCD, the latter is not so widely used in practice, where lattices tend to be more elongated in time direction.

For the purpose of illustrating discretization and finite-volume errors, we have considered four cases:
\begin{itemize}
	\item[i)] continuum ($x$ takes any real value)
	\item[ii)] a discrete lattice in position space with $x=a n$ and $n$ any integer number (note that this case considers an infinite volume), 
	\item[iii)] a finite box of size $L$ with periodic boundary conditions with $x$ taking any value in the range $(0,L)$, and 
	\item[iv)] the lattice case, with $x$ discrete and belonging to a box with periodic boundary conditions.
\end{itemize} 
For latter convenience, it is important to remark that the possible values of the momenta in each one of the cases mentioned above are determined by the restrictions in the positions. In this sense, if each component of the position takes the discrete form  $x_i=a n_i$, where $a$ is the {\it lattice spacing} and $n_i$ an integer, the accessible momenta will be restricted to the range $(0,2\pi/a)$, while if the positions are restricted to be in a range $(0,L)$, the accessible momenta will be discrete and of the form $\frac{2\pi}{L} n_i$. Those relations between positions and momenta have been detailed in table ~\ref{tab:i}

\begin{table}[!h]
	\centering
	\begin{tabular}{|l|c|c|}
		\hline
		Case & position & momenta \\
		\hline 
        \textit{i)} Continuum, infinite volume & $x_\mu \in R $ & $k_\mu\in R$ \\
		\hline
        \textit{ii)} Discrete, infinite volume & $x_\mu=a\ n_i$ & $k_\mu\in \left(0,\frac{2\pi}{a}\right)$ \\
        & $n_\mu\in \text{Integers}$ & p.b.c. \\
\hline
        \textit{iii)} Continuum, finite volume & $x_\mu=\in (0,L)$ & $k_\mu=\frac{2\pi}{L} n_\mu$ \\
        & p.b.c. & $n_\mu \in \text{Integers}$ \\
\hline
        \textit{iv)} Discrete, finite volume  & $x_\mu=a\ n_\mu$ & $k_\mu=\frac{2\pi}{Na}  n_\mu$ \\
            (lattice)                & $n_\mu=0,1,\cdots,N-1$ & $n_\mu=0,1,\cdots,N-1$ \\
\hline
	\end{tabular}
	\caption{\label{tab:i} Possible values of each component of positions and momenta for each one of the four cases considered.}
\end{table}

\subsection{Klein-Gordon propagator}

\paragraph{Momentum space.}
The propagator of a free boson field of mass $\mu$ in a continuum, infinite space (case \textit{i)} in Tab.~\ref{tab:i}) is given by:
\beq\label{KGcontinuumP}
\Delta_\text{cont}(k) = \frac{1}{k^2+\mu^2}\ .
\eeq
with $k^2=\sum_{\mu=1}^4 k_\mu^2$.
The same expression holds when working in a finite box with periodic boundary conditions (case \textit{iii)} in Tab.\ref{tab:i}), with the particularity that in this case only discrete values of the momenta are allowed. 

If a discretization is introduced in position space so that the allowed positions take the form $x_i=a n_i$, the propagator has the form: 
\beq\label{KGdiscreteP}
\Delta(k) = \frac{1}{\hat{k}^2 + \mu^2}\ ,
\eeq
where $\hat{k}^2=\sum_{\mu=1}^4 \hat{k}_\mu^2$ and $\hat{k}_\mu$ are the lattice momenta that, using the forward discretization of derivatives (\ref{eq:forward}), take the form: $\hat{k}_\mu=\frac{2}{a} \sin \frac{ak_\mu}{2}$. Note that this expression holds for discrete positions, both in the case they span over an infinite volume or are limited to a finite lattice (cases \textit{ii)} and \textit{iv)} in Tab.~\ref{tab:i} respectively). The only difference between those two cases is that in the former $k_i$ can take any real value in the range $(0,2\pi/a)$ while in the latter the allowed momenta are discrete with  $k_\mu=\frac{2\pi}{N a} n_\mu$ with $n_\mu=0,1,\cdots,N-1$.

It is noticeable that the presence of a finite volume does not affect the continuum propagator  (\ref{KGcontinuumP}). This means that a free boson propagator in a finite volume will be identical --for the momenta where it can be evaluated-- to the infinite volume result. This is no longer true in an interacting theory, where the interactions can introduce a volume dependence. In any case, the fact that finite volume effects do not affect the tree-level propagator may be at the origin of the subdominant role played by finite volume effects in gluon propagator. It is now widely accepted that finite-volume effects are almost negligible~\cite{Boucaud:2011ug} above $\sim 0.5{\rm GeV}$ but they become an important issue as the infrared limit is approached~\cite{Oliveira:2005hg,Bogolubsky:2007bw,Cucchieri:2008mv,Oliveira:2009nn,Bornyakov:2009ug}. The modern capability of producing very large volume lattice simulations with a lattice size $L\gtrsim 10{\rm fm}$ significatively reduces even more the impact of finite volume errors~\cite{Boucaud:2018xup,Aguilar:2019uob}. 
For the particular case of gluon propagator, taking the infinite volume limit for the space of local gauge transformations~\cite{Zwanziger:1993dh,Cucchieri:2016qyc} has proven to be an efficient method to significantly reduce finite volume errors affecting the low momenta. This would mean that finite volume effects for the deep IR of gluon propagator would be associated to the Landau gauge-fixing in the lattice and can be therefore cured by replicating the finite lattice and fixing the gauge over the extended configuration.

\paragraph{Position space.}

The position space propagator can be obtained from the momentum space one via Fourier transform. In the case of a four-dimensional continuum, infinite space it is given by:
\beq\label{KGcontinuumX}
\Delta_\text{cont}(x) = \int\ \frac{d^4p}{(2\pi)^4} e^{- i p x} \Delta_\text{Cont}(p) = 
 \frac{1}{4 \pi^2 x} \int_0^\infty dp\ p^2 \Delta_\text{Cont}(p) J_1(p x) = \frac{\mu\ K_1 (\mu\ x)}{4 \pi^2 x}\ ,
\eeq
where we used the  rotational symmetry for integrating out the angular part and $J_n(z)$ and $K_n(z)$ represent Bessel functions of the first kind and modified second kind respectively.

\medskip

In an box of finite volume, the position space propagator is given by: 

\beq\label{KGfiniteX}
\Delta(x) = \frac{1}{L^4} \sum_k {\Delta}(k) e^{-i k x} 
\eeq
where 
\begin{itemize}
	\item for the continuum case (labeled \textit{iii)} in Tab.~\ref{tab:i}) the sum is extended to an infinite number of momenta in the four dimensions of space-time and the ${\Delta}(k)$ corresponds to Eq.~(\ref{KGcontinuumP}).
	\item for the discrete case (labeled \textit{iv)} in Tab.~\ref{tab:i})  the sum is extended to the limited set of accesible momenta, $k_\mu=\frac{2\pi}{Na} n_\mu$, and ${\Delta}(k)$ corresponds to Eq.~(\ref{KGdiscreteP}).
\end{itemize}

Finally, for the infinite-volume discrete case (labeled \textit{ii)} in Tab.\ref{tab:i}) the values of the propagator in position-space are 
obtained through:
\beq\label{KGdiscreteX}
\Delta(x) = \int_{k_\mu\in\left(0,\frac{2\pi}{a}\right)} d^4k\  \Delta(k) e^{-i k x}
\eeq
where $\Delta(k)$ is given by Eq.(\ref{KGdiscreteP}).

From their construction, the deviations of Eq.~(\ref{KGfiniteX}) for the continuum case with respect to Eq.~(\ref{KGcontinuumX}) will be \textit{finite-volume errors} while the deviations of Eq.~(\ref{KGdiscreteX}) with respect to Eq.~(\ref{KGcontinuumX}) will be \textit{discretization errors}. The lattice propagator obtained using the discrete momentum space propagator (\ref{KGdiscreteP}) in (\ref{KGfiniteX}) will include a mixture of both discretization and finite-volume errors and constitutes the actual object of study of this paper.

\subsection{Rotational symmetry breaking artifacts in Klein-Gordon propagator}

\paragraph{Momentum space\\}

\begin{figure}[ht]
	\begin{center}
		\begin{tabular}{c c}
			\includegraphics[width=0.5\textwidth]{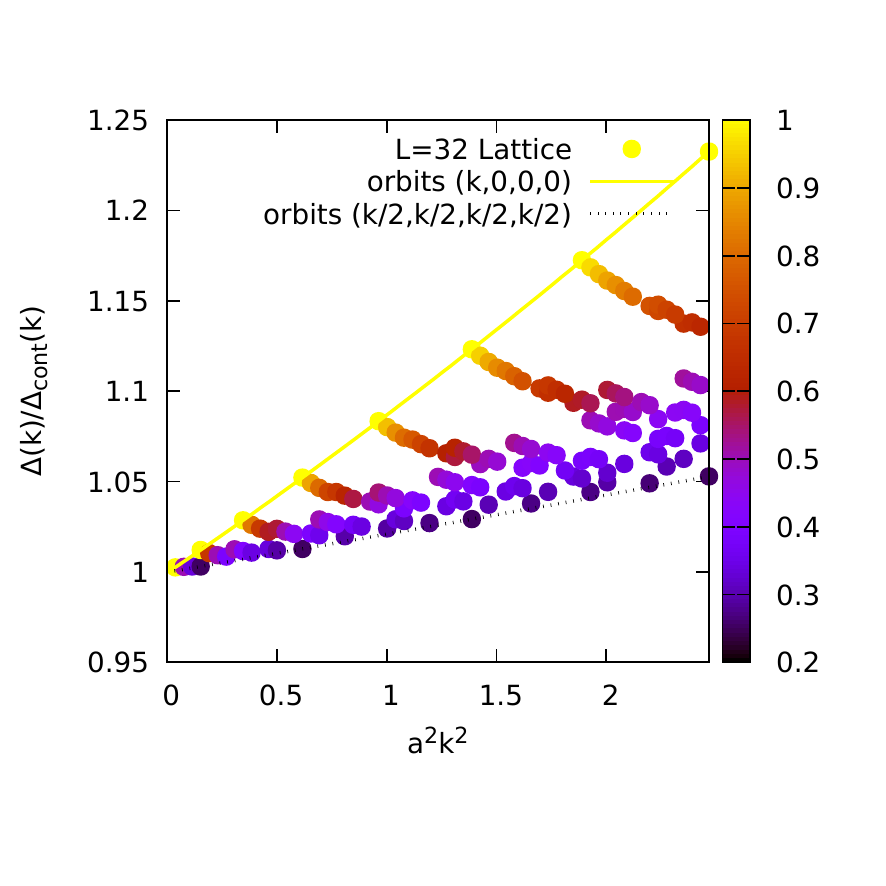} &
			\includegraphics[width=0.5\textwidth]{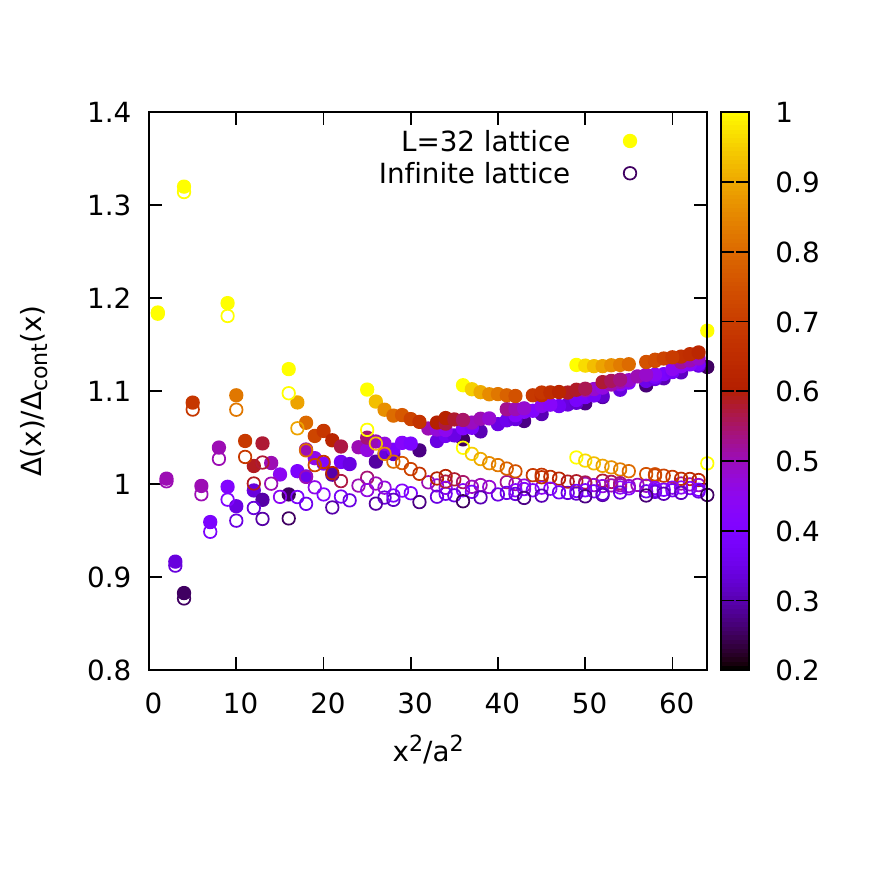} \\
			(a) & (b)
		\end{tabular}
	\end{center}\caption{Ratio of the lattice Klein-Gordon propagator for $N=32$ and $a\mu=0.1$ and its continuum counterpart in momentum (a) and position (b) space. In (a) full lines have been added for the two limiting cases of orbit along an axis and along the diagonal in a continuum box. In (b), empty symbols represent the discrete infinite volume case. In both plots, the color map corresponds to the dimensionless ratio $n^{[4]}/(n^2\cdot n^2)$.}
	\label{fig:dataXP}
\end{figure}

In order to illustrate the role of lattice artifacts associated to the rotational symmetry breaking, we have plotted in Fig.\ref{fig:dataXP}(a) the ratio
\begin{eqnarray}
\frac{{\Delta}(k)}{{\Delta}_{\rm cont}(k)}
\end{eqnarray}
between the momentum-space Klein-Gordon propagator in a lattice (\ref{KGdiscreteP}) and its continuum counterpart (\ref{KGcontinuumP}) for a lattice of size $N=32$ and a boson mass~\footnote{From now on, all dimensional quantities are expressed in terms of $a$. The value $a\mu=0.1$ chosen would correspond to a boson of mass $\mu\sim 0.4{\rm GeV}$ for a lattice spacing $a\sim 0.05{\rm fm}$. Note that those values are in the approximate range of the dynamically generated gluon mass~\cite{Aguilar:2019uob} and typical values of spacings in current lattice QCD calculations.} $a\mu=0.1$. As this ratio should be $1$ in the continuum limit, deviations from $1$ will be discretization errors.
In the plot, each momentum $(k_1,k_2,k_3,k_4)$ is represented separately. One observes that the plot not only deviates from $1$, but some values of the squared momenta, $k^2$, there are multiple values of the propagator.
This is due to the fact that momenta not related by the discrete lattice group  H(4), i.e., by permutations among the components of $k$ or sign changes, give a different result. We will refer to each set of momenta related by the H(4) symmetry group as orbit. Thus, instead of obtaining a smooth function that only depends on $k^2$ as in the continuum, one obtains a different value of the propagator for each orbit. Indeed, it depends not only on $k^2$ but also on the higher order invariants of the H(4) group: $k^{[4]}=\sum_\mu k_\mu^4$, $k^{[6]}=\sum_\mu k_\mu^6$, and $k^{[8]}=\sum_\mu k_\mu^8$. To illustrate this, we picked the invariant $k^{[4]}$, which will have the leading role, and used the dimensionless ratio:
\beq\label{ratio4}
\frac{k^{[4]}}{(k^2)^2}
\eeq
to set the color scale of Fig.\ref{fig:dataXP}(a). Note that $\frac{k^{[4]}}{(k^2)^2}$ takes values ranging from $1/4$ for momenta distributed along the lattice diagonal, i.e. as $(\frac k 2,\frac k 2,\frac k 2,\frac k 2)$ to $1$ for momenta along an axis, $(k,0,0,0)$.

At first sight, discretization errors are highly correlated with the value of $\frac{k^{[4]}}{(k^2)^2}$, being the orbits whose momentum is maximally shared among the four directions (called {\it democratic} orbits) closest to the continuum result and the orbits with the momentum along an axis the most distant ones.
The deviations from $1$ in Fig.~\ref{fig:dataXP}(a) are responsible for the well-known half-fishbone structures that tend to appear in momentum-space propagators~\cite{Becirevic:1999hj,Boucaud:2011ug}. For comparison, we have also plotted the lines for the momentum-space propagator in Eq.~(\ref{KGdiscreteP}) for an infinite-volume (where each component of the momenta can take any real value), for momenta with extreme values of the ratio in Eq.~(\ref{ratio4}), i.e. along the lattice diagonal and axis.

This picture is, of course, very well known, and has triggered different methods to recover rotationally invariant results. The role of the H(4) invariants in the appearance of these discretization errors can be visualized from a Taylor series  of the discrete propagator (\ref{KGdiscreteP}) in the lattice spacing $a$, 
\begin{equation}\label{eq:taylor}
	\Delta(k) = \frac{1}{\hat{k}^2 + \mu^2} = \frac{1}{k^2 + \mu^2} + \frac{a^2}{12} \left( \frac{1}{k^2 + \mu^2} \right)^2 k^{[4]} + ...
	\ ,
\end{equation}
whose first term is proportional to the H(4) invariant $k^{[4]}$ (see \cite{deSoto:2007ht} for more details). The fact that momenta along the lattice diagonal have the minimum~\footnote{Recall that democratic orbits $(\frac{k}{2},\frac{k}{2},\frac{k}{2},\frac{k}{2})$ have the minimum $k^{[4]}$ among all the orbits with the same $k^2$.} contribution to the leading $a^2$ correction in  (\ref{eq:taylor})  are the origin of the {\it democratic} methods~\cite{Leinweber:1998uu} where only orbits along or close to the diagonal are considered (the rest is discarded). H4 extrapolation methods~\cite{deSoto:2007ht} go one step forward and use all the lattice orbits to extrapolate the higher order invariants to zero. A different and very successful course of action~\cite{Bonnet:2001uh} is plotting the lattice data as a function of $\hat{k}^2$ instead of $k^2$ and use a correction to the lattice propagator based in the lattice tree-level propagator, ${\Delta}_{\rm TL}(k)$:
\begin{eqnarray}\label{eq:TLcorr}
	{\Delta}_{\rm Corr}(k) = {\Delta}_{\rm Latt}(k) \frac{{\Delta}_{\rm Cont}(k)}{ {\Delta}_{\rm TL}(k)}\ .
\end{eqnarray}
For the free case studied in this section, this method is exact, and the use of any of those two conditions separately would provide the continuum result~\footnote{Using $\hat{k}$ already solves the problem as (\ref{KGdiscreteP}) is the same function than (\ref{KGcontinuumP}) changing $k$ by $\hat{k}$. On the other hand, as the lattice result is tree-level, the ratio in (\ref{eq:TLcorr}) equals the continuum result by definition.}.

\paragraph{Position space\\} 

In analogy with what has been done in momentum space, the ratio of the Klein-Gordon propagator in  position-space in a lattice (\ref{KGdiscreteX}) to the continuum one (\ref{KGcontinuumX}) has been plotted in Fig.~\ref{fig:dataXP}(b) for the same parameters. The color scale indicates in this case the value of the dimensionless ratio:
\beq\label{ratio4x}
\frac{x^{[4]}}{(x^2)^2}\ .
\eeq
The plot includes both the lattice propagator for a finite volume (full symbols) and an infinite lattice (empty symbols). Both sets show a large spreading of the data for small distances, while for large $x$ they tend to concentrate along a curve (around one for the infinite volume case and away from one for the finite-volume case). It is important to remark that the difference between both sets is a finite-volume effect which is responsible for the  differences at large distances. For short distances, nevertheless, the data are almost identical in the two sets signaling the presence of rotational symmetry breaking discretization artifacts that are large for this region. 

If one considers the infinite-volume data (to avoid mixing in our discussion finite-volume and discretization errors) of Fig.~\ref{fig:dataXP}(b), one can observe that democratic orbits tend to be below $1$ in the plot, and therefore, the method based in a cylindrical cut along the diagonal will not work in general in position space. Indeed, a closer look to the data shows that in this case the ``best'' orbits (in the sense they are closer to the continuum result) have $x^{[4]}/(x^2)^2\sim 1/2$.

Contrarily to the momentum-space case, discretization and finite-volume artifacts are not well known in position-space and, up to the knowledge of the author, the only method that has been applied in position space~\cite{PhysRevLett.125.242002} is a tree-level correction as the one sketched in Eq.~(\ref{eq:TLcorr}) for momentum-space. Among the reasons that may explain why discretization errors are harder to eliminate in position space, one can signal the fact that while in momentum space, discretization errors have a sizable effect for large values of $k^2$, where there are many lattice orbits $(k_1,k_2,k_2,k_4)$ with the same $k^2$, in position space discretization errors affect mainly the small-x region, where few orbits are available. This asymmetry explains why extrapolations in $k^{[4]}$ efficiently remove discretization errors while in $x^{[4]}$ the results are rather poor. Moreover, while in momentum space \textit{democratic} orbits can be used to reduce (at least partially) discretization errors, it is no longer true in position-space, where \textit{democratic} orbits for small-$x$, where it exhibits large deviations from the continuum value (see Fig.~\ref{fig:dataXP}(b)). In next section we will propose a method to deal with discretization (and possibly finite-volume) errors in an unified way for both momentum and position-space.

\section{Generalized H4 method}
\label{sec:cure}

The observation~\cite{Becirevic:1999hj,Becirevic:1999uc} of a linear trend in $k^{[4]}$ in the lattice data for any fixed value of $k^2$ motivated the introduction of the so-called H4 extrapolation methods, that use the different orbits with the same $k^2$ to make an extrapolation of $k^{[4]}$ to zero. This fit, according to Eq.(\ref{eq:taylor}), would remove the leading discretization errors of order  $\mathcal{O}(a^2)$ for a free massive boson propagator. In this paper we generalize and extend the  method originally developed in \cite{Becirevic:1999uc,Becirevic:1999hj,Boucaud:2003dx} and described in detail in \cite{deSoto:2007ht}.

H4-extrapolation methods start by assuming an expansion~\footnote{For clarity only the leading term in the invariant $k^{[4]}$ has been included here, although higher order terms can be included, as detailed in \cite{deSoto:2007ht}. A more recent application of the method that takes into account $k^{[6]}$ and ${k^{[4]}}\cdot{k^{[4]}}$ terms can be found in \cite{Oliveira:2018lln}.} inspired by the Taylor series in (\ref{eq:taylor}:)
\beq\label{H4classic}
F(k^2)= F(k^2,k^{[4]}) + a^2 \ C(k^2) \ k^{[4]}
\eeq
where $C(k^2)$ is some function of $k^2$ that is fitted either separately for each $k^2$ or --assuming it is a smooth function of $k^2$-- for a range of values of $k^2$. This method has been successfully used over the last two decades for eliminating H4 errors from lattice gluon and ghost propagators~\cite{Boucaud:2011ug,Boucaud:2018xup,Aguilar:2019uob}. The role played by higher order corrections (such as $k^{[6]}$, $(k^{[4]})^2$, etc) is particularly relevant for large momenta and it is therefore dangerous to extent the H4-extrapolation to very large momenta without adding more terms. In general, we limited the range of momenta to $k^2_{\max} < \left(\frac{2\pi}{Na}\right)^2 n_{\max}^2$, with $n_{\max}=N/4$. A noticeable effort in the extension of this method for the understanding of the structure of lattice gluon propagator and its discretization artifacts using a different point of view has been presented in \cite{Catumba:2021hcx,Catumba:2021pdu}.

\medskip

In order to generalize the method for both positions and momenta, let us re-write this expression i) in terms of $n^2$, $n^{[4]}$ which is a notation valid both for momentum-space where $k^2=(\frac{2\pi}{L})^2\ n^2$ and position-space with $x^2=a^2\ n^2$, ii) using the dimensionless ratio in Eqs.~(\ref{ratio4}-\ref{ratio4x}) --note that this ratio is dimensionless both in position and momentum space and can be written as $n^{[4]}/{(n^2)}^2$, iii) as a multiplicative correction, which will allow a dimensional analysis exercise and is flexible to be applied to a wide variety of functions, and iv) adding a parameter $\alpha_4$ to account for the fact that the $n^{[4]}\to 0$ limit does not necessarily recover the continuum results (as it is observed in position space, remind Fig.~\ref{fig:dataXP}b). With all, it results:
\beq\label{H4new}
F(n^2)= F(n^2,n^{[4]}) \times \left[ 1 + c(n^2)  \left(\frac{n^{[4]}}{(n^2)^2}-\alpha_4 \right)\right]
\eeq
Note that Eq.~(\ref{H4classic}) is recovered from Eq.~(\ref{H4new}) with $\alpha_4=0$ and $a^2 C(p^2) \propto F(n^2) \times \frac{c(n^2)}{(n^2)^2}$. In this work, instead of fitting $c(n^2)$ for each $n^2$ as an independent parameter, we will make a global fit for all the values of $n^2$, searching the optimal functional form of $c(n^2)$. It is important to remark that this is still only the first correction, and therefore it should not be extended to too large values of $n^2$ where higher order invariants such as $n^{[6]}$ may play an important role.

The addition of the $\alpha_4$ term in Eq.~(\ref{H4new}) is not an H4-breaking term but rather it prevents the H4 extrapolation to introduce O4 errors, which would be the case in position-space as can be observed in Fig.~\ref{fig:dataXP}(b) and will be discussed below. It is important to remark that eliminating H4 errors produce smooth results (in contrast with the noisy fluctuations obtained otherwise) but not necessarily eliminates discretization errors. For illustration, consider democratic orbits in Fig.~\ref{fig:dataXP}(a); they have a smooth behavior in $k^2$ but it is clearly different from one, and indeed shows a linear trend with $k^2$.

\paragraph{Application to momentum-space \\ }

Let us consider the data for the lattice boson propagator in momentum space and apply Eq.~(\ref{H4new}). We take the lattice data $F(n^2,n^{[4]})$ for $n^2<(N/4)^2$ and fit  the artifact-free value $F(n^2)$ and the function $c(n^2)$ using $\alpha_4=0$. 
To avoid any \textit{a priori} guess about $c(n^2)$, we made a fit~\footnote{For the purpose of these fits, we have assigned an arbitrary statistical error of $0.1\%$ to all the exact lattice data used. Note that this value is not unrealistic in current lattice simulations with a large number of gauge field configurations.} using $c(n^2)=c_m (n^2)^m$, with $m$ an integer number and $c_m$ a free parameter. The result obtained is that the minimum $\chi^2$ is obtained for $m=1$. Indeed, we could have anticipated this result using a simple dimensional analysis, as the simplest way of obtaining a $k^2$-dependent term of order $a^2$ is $a^2 k^2$ which implies for the dimensionless function $c(n^2)\propto n^2$.  

Once we know the fit prefers positive powers of $n^2$, we finally use a more general  form such as: 
\beq\label{cP}
c(n^2) = c_0 + c_1 n^2 + c_2 (n^2)^2
\eeq
and the parameters $c_0$, $c_1$, $c_2$ as well as the extrapolated values $F(n^2)$ for each $n^2$ are the free parameters of the fit. Note that the number of orbits used (i.e., the number of input values $F_L(n^2,n^{[4]})$) is much larger than the number of fitted quantities ($n^2+3$ in this case). As an example with the four dimensional lattice of $N=32$ points in each direction used, and keeping only orbits up to $n^2=(L/4)^2$ we have $157$ orbits for $64$ values of $n^2$ and the three parameters $c_0$, $c_1$, and $c_2$. It is important to remark that the number of orbits grows very fast with $n^2$, and faster for larger space-time dimensions.

\begin{figure}[h]
	\begin{center}
    \begin{tabular}{c c}
		\includegraphics[width=0.5\textwidth]{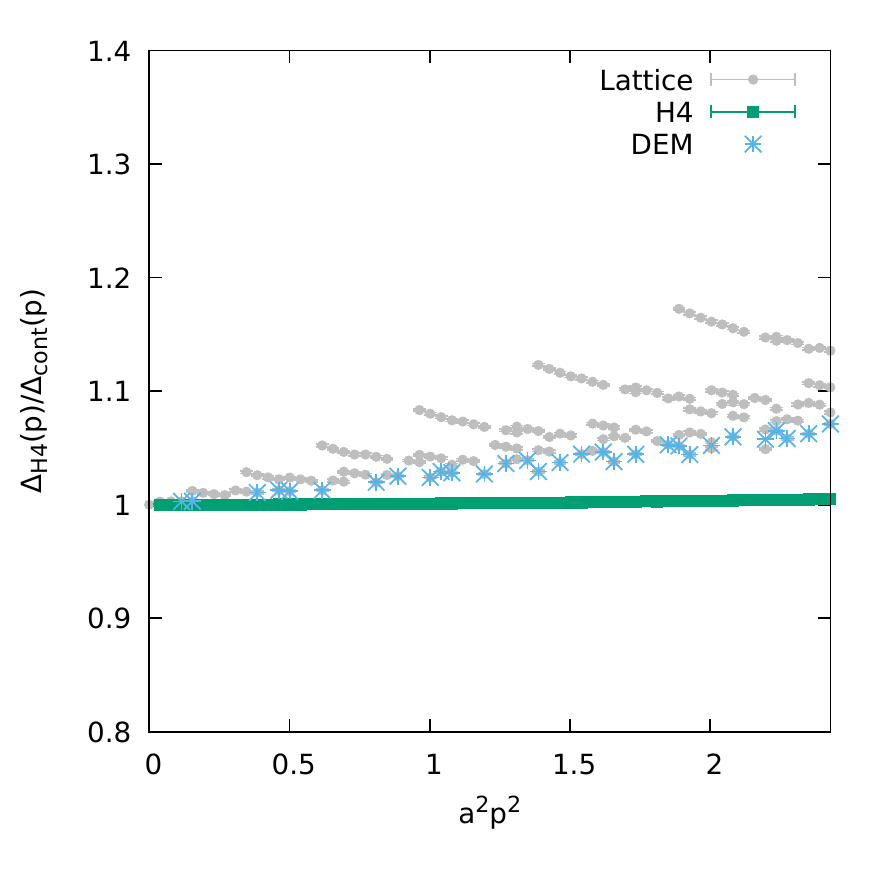} &
		\includegraphics[width=0.5\textwidth]{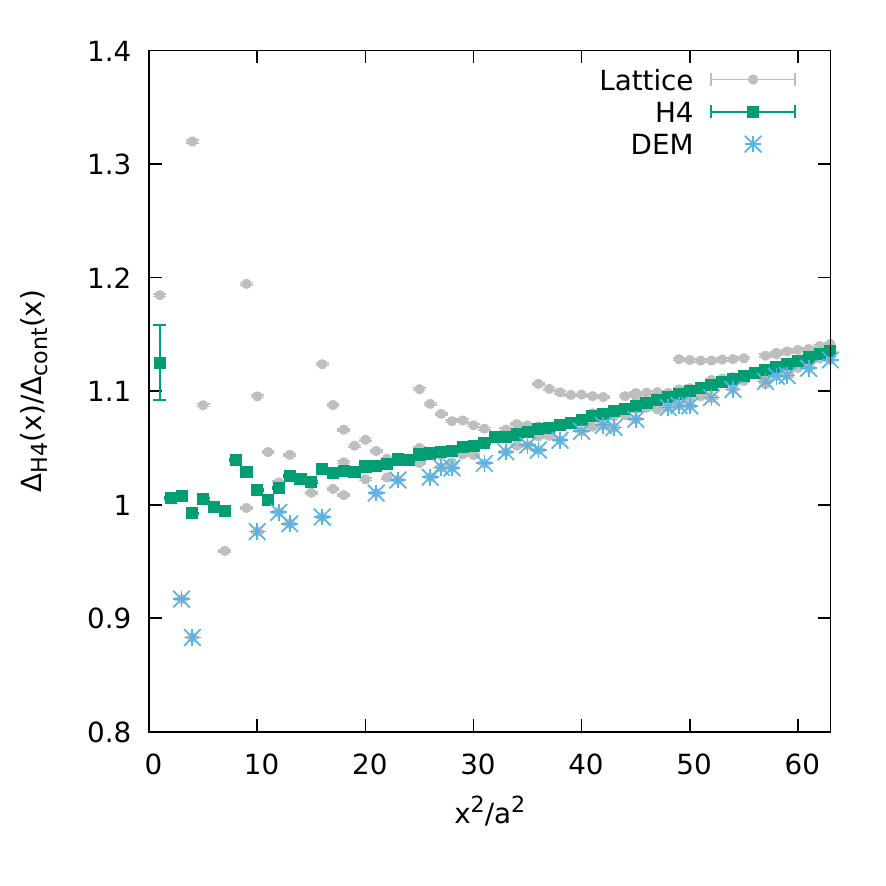} \\
		(a)&(b)
	\end{tabular}
	\end{center}\caption{Ratio of Klein-Gordon propagator in a $32^4$ lattice  with respect to the continuum one in momentum space (a) and position-space (b). Grey circles represent the original lattice data and green squares the result of applying the generalized H4 method. For comparison, the results of a democratic cut with an angle smaller than $30^\circ$ with diagonal have been added as light blue stars.}
	\label{fig:extra}
\end{figure}

Fitting the exact propagator in Eq.~(\ref{KGdiscreteP}) with the form (\ref{H4new}) and using $\alpha_4=0$ and the expansion (\ref{cP}), we obtain the coefficients: $c_0=(0.5\pm 1.6)\cdot 10^{-3}$, $c_1=(-3.2\pm 0.1)\cdot 10^{-3}$ and $c_2=(4\pm 1)\cdot 10^{-6}$. With the results of the fit we have plotted in Fig.~\ref{fig:extra}(a) the ratio of  H4 extrapolated propagator to the continuum one. In the plot one can observe that the deviation from the exact continuum propagator after the H4 extrapolation is very small (below $1\%$), while using the democratic method it would have a already a bias of $\sim 5\%$ for the end of our fitting window. 

Notice that the value obtained in the fit for the dominant contribution, $c_1$, is compatible with the one that can be deduced from Eq.(11) of Ref.~\cite{deSoto:2007ht}, which would be $-\frac{1}{12}\left(\frac{2\pi}{N}\right)^2\approx -3.213\cdot 10^{-3}$ for $N=32$.

\paragraph{Application to position-space \\ }

Let us now apply the method explained above to the position-space data represented in Fig.~\ref{fig:dataXP}(b). We already discussed above that breaking the rotational symmetry has a larger impact for small distances (small $n^2$). This affects the capabilities of the traditional H4-extrapolation methods for eliminating the artifacts because in this region there is a small number of orbits. An additional difficulty is that the ``optimal'' value of the ratio (\ref{ratio4x}) seems to be $1/2$ instead of $0$ in this case. Let us fix $\alpha_4=1/2$ based in this empirical observation (we will come back to this issue later) and perform the same analysis we did in momentum-space.  The minimum $\chi^2$-fit is obtained in this case for $c(n^2)\sim 1/n^2$, which again can be inferred from dimensional analysis, as a $x^2$-dependent artifact of order $(a^2)$ is of the form $a^2/x^2\sim (n^2)^{-1}$.

The result of the H4 extrapolation~\footnote{To maintain the name of the method, we use here the term ``extrapolation'' although in this case the $\alpha_4=1/2$ lies in the middle of the lattice orbits.} is represented in Fig.~\ref{fig:extra}(b) using in this case
\beq\label{cX}
c(n^2) = c_0 + \frac{c_{-1}}{n^2} + \frac{c_{-2}}{(n^2)^2}
\eeq
with the best-fit values $c_0=(3.1\pm 0.9)\cdot 10^{-3}$, $c_{-1}=-2.71\pm 0.03$, and $c_{-2}=2.8 \pm 0.1$. We checked that adding more terms to the development in (\ref{cX}) does not significantly increase the quality of the fit.

Although the magnitude of the errors in the data is highly reduced in this case as well, there are two features in Fig.~\ref{fig:extra}(b) which contrast with the situation in the momentum-space case of Fig.~\ref{fig:extra}(a). First, there is a non-vanishing slope in $x^2$ in the data after applying the H4 extrapolation. This signals the presence of finite volume errors that grow with $(x/L)^2$ and that are not present in the infinite-volume discrete data of Fig.~\ref{fig:dataXP}(b). Second, there is still a noticeable noise in the data for small $x^2$ (mainly below $(x/a)^2\sim 15$). This problem seems to be related to the fact that the region with larger discretization errors is also the one with a smaller number of orbits, as discussed above. 

In our application of the generalised H4 extrapolation method both in momentum and position space we focused in minimizing $\chi^2$, i.e., obtaining the best functional description of the data using Eq.(\ref{H4new}). The method  highly reduces the noise associated to the symmetry breaking artifacts and one could add an extra condition over the smoothness of the extrapolated results. Although it could provide an useful guide, we preferred here not to impose smoothness, as it might add uncontrolled biases to the results.

\section{Continuous Fourier transforms and O4-invariant artifacts}  
\label{sec:ft}

Even if the noise associated with rotational breaking artifacts had been completely removed, the smooth results obtained could still be affected by O4-symmetric discretization errors. In some cases, the method used to recover rotational invariance may introduce an O4-symmetric artifact.
Nevertheless, the propagators in the continuum infinite-volume limit in position and momentum space (let us call them generically $F(x)$ and ${F}(k)$) must be one the Fourier transform of the other, i.e.:
\begin{eqnarray}
{F}(k) &=& \int d^4x\ e^{-i k x} F(x) = \frac{4\pi^2}{k}  \int_0^\infty dx\ x^2\ J_1(k x) F(x) \quad \Leftrightarrow \label{ftk} \\
F(x) &=&             \int \frac{d^4k}{(2\pi)^4} e^{i k x} {F}(k) = \frac{1}{4\pi^2 x}  \int_0^\infty dk\ k^2\ J_1(kx) {F}(k) \label{ftx} 
\end{eqnarray}
where in the second equality the angular part has been integrated out.

Once we have applied the generalized H4 method introduced in previous section, we can use the expressions (\ref{ftk}) and (\ref{ftx}) to check whether the functions obtained satisfy those equations. If we had correctly recovered the continuum- infinite-volume results through the suppression of the spurious dependence in the higher order invariants of the H4 group, the Fourier transform of the extrapolated result in position space would coincide with the extrapolated result in momentum-space and vice-versa. 

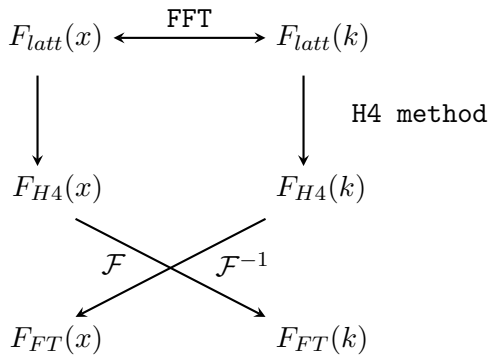
\begin{figure}[h]
	\begin{center}
		\begin{tikzpicture}[scale=1]
			\draw [black, thick, <->, >=stealth] (1.,5.) -- (3,5) node [midway, above] {\textcolor{black}{\texttt{FFT}}};
			\draw (1,5) node[left] {\textcolor{black}{$F_{latt}(x)$}};
			\draw (3,5) node[right] {\textcolor{black}{${F}_{latt}(k)$}};
			\draw [black, thick, ->, >=stealth] (0.,4.5) -- (0,3.3);
			\draw [black, thick, ->, >=stealth] (3.5,4.5) -- (3.5,3.3);
			\draw (4,4) node[right] {\textcolor{black}{\texttt{H4 method}}};
			\draw (1,3) node[left] {\textcolor{black}{$F_{H4}(x)$}};
			\draw (3,3) node[right] {\textcolor{black}{${F}_{H4}(k)$}};
			
			\draw [black, thick, ->, >=stealth] (0.5,2.6) -- (3,1.3);
			\draw [black, thick, ->, >=stealth] (3,2.6) -- (0.5,1.3);
			
			\draw (2.7,2) node {\textcolor{black}{$\mathcal{F}^{-1}$}};
			\draw (1.,2) node {\textcolor{black}{$\mathcal{F}$}};
			
			\draw (1,1) node[left] {\textcolor{black}{$F_{FT}(x)$}};
			\draw (3,1) node[right] {\textcolor{black}{${F}_{FT}(k)$}};
			
			
		\end{tikzpicture}
	\end{center}
    \caption{Schematic representation of the definition of propagators via Fourier transforms in the continuum of the H4-extrapolated data. If the method removed all lattice artifacts (both discretization and finite volume ones), $F_{FT}(x)$ and $F_{H4}(x)$ would coincide, and the same can be said about momentum-space propagators.}
\end{figure}

Computing the Fourier transform of the propagators that we know only at some discrete values of the distance or momentum (those corresponding to an integer value of $n^2$) is not straightforward. In order to compute the integrals in Eqs.(\ref{ftk}-\ref{ftx}), we need a continuous description of $F(x)$
for all values of $x$ between $0$ and $\infty$ (and respectively ${F}(k)$ for all values of $k$). As we have limited information for integer values of $n$, we are forced to make some approximations. In the present work we have done the following:
\begin{itemize}
	\item For small values of $n^2$ we use a Padé approximant 
	\beq
	\frac{a_0+a_1 n^2 + a_2 (n^2)^2}{1+b_1 n^2 + b_2 (n^2)^2}
	\eeq
	whose coefficients are by obtained from a fit  $n^2<n^2_{\text min}$ in each case. This behavior is assumed to describe the continuous function between $n^2=0$ and $n^2_{\text min}$. We fixed $n^2_{\text min}=6$ for this work.
	\item Among any two consecutive values of $n^2$ we  assume a linear interpolation between the lattice points.
	\item For values of $n^2$ larger than the maximum value of $n^2$ considered, $n^2_{\text max}=(N/4)^2$, we extend the propagator with the functional form:
	\beq
	\frac{d_1}{n^2} +  	\frac{d_2}{(n^2)^2} +  	\frac{d_3}{(n^2)^3} 
	\eeq
    where the  coefficients $d_i$ are fitted over the largest values of $n^2$ accessible. 
\end{itemize}
Once we have set a continuous description of the discrete data, we perform numerically the integrals in Eqs.(\ref{ftk}-\ref{ftx}) to obtain an approximation to the Fourier transforms in the continuum. 
Although there is an unavoidable bias due to our lack of information (in particular for the limits $n^2\to 0$ and $n^2\to\infty$), this procedure allows us to make a comparison between the H4-extrapolated propagator $\Delta_{H4}(x)$ and the one obtained through Fourier transform of ${\Delta}_{H4}(k)$, ${\Delta}_{FT}(x)$.

\medskip

\begin{figure}[h]
	\begin{center}
	\begin{tabular}{c c}
		\includegraphics[width=0.49\textwidth]{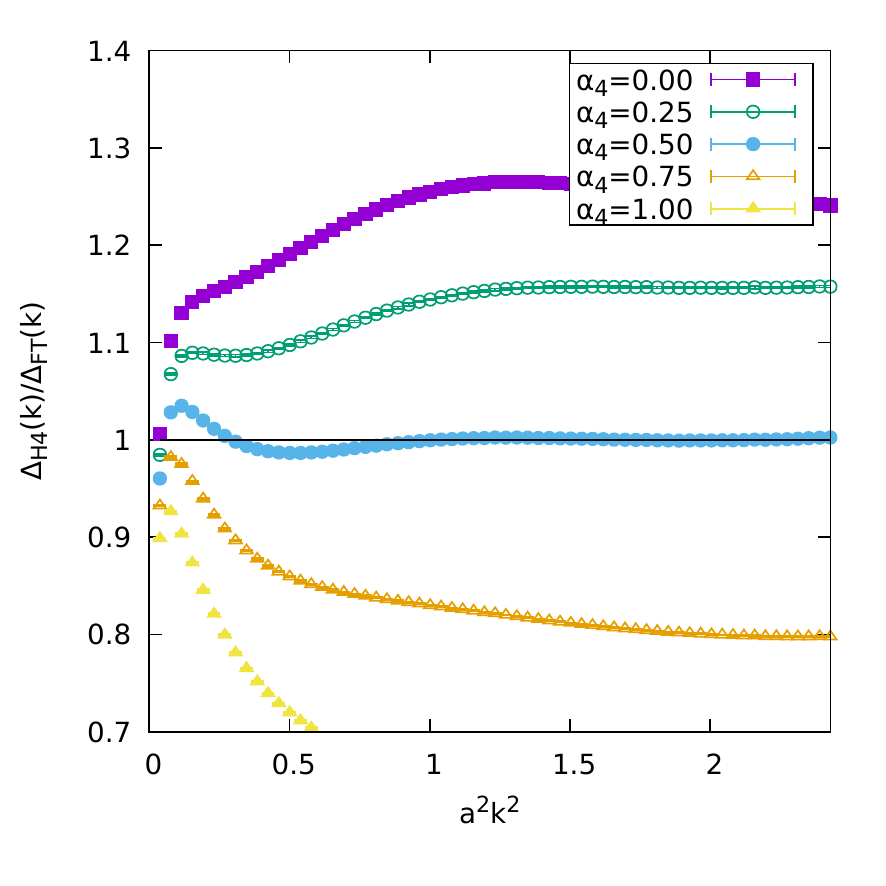} &
		\includegraphics[width=0.49\textwidth]{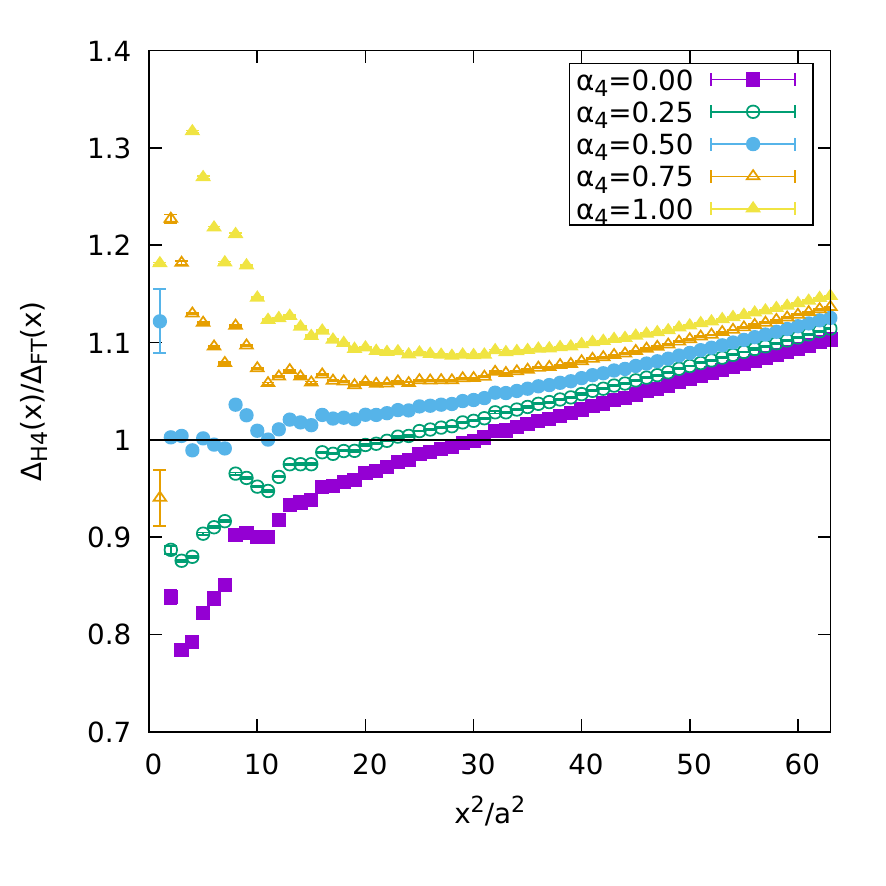}
    \end{tabular}
	\end{center}\caption{Ratios of the Klein-Gordon propagator obtained through the generalized H4 method and the one coming from the approximate continuum Fourier transform. Left plot corresponds to momentum-space and right on to position-space.
	In both plots, each set corresponds to a different value of the parameter $\alpha_4$.}
	\label{fig:ratioXP}
\end{figure}

In order to show the results of the Fourier transform, we will use the ratio of the H4-extrapolated propagator in momentum space, ${\Delta}_{H4}(k)$, divided by
the Fourier transform of the H4-extrapolated one in position space,
\beq
{\Delta}_{FT}(k) = \frac{4\pi^2}{k}  \int_0^\infty dx\ x^2\ J_1(k x) \Delta_{H4}(x) .
\eeq
This ratio would be $1$ if we had fully eliminated lattice errors and if Fourier transform was exact.
We have represented this ratio for the Klein-Gordon case studied in the previous section in Fig.~\ref{fig:ratioXP}(left) for five different values of the parameter $\alpha_4$ used in the position-space H4 extrapolation ranging from $\alpha_4=0$ to $\alpha_4=1$. The figure shows that except in the case $\alpha_4=0.5$, the ratio deviates significantly from $1$, providing  a sound argument for the choice of $\alpha_4=0.5$ we made. 
Let us remark that within this plot we did not use the exact solution in the continuum infinite-volume for this case, but only the results of the H4 extrapolation of the lattice data in both position and momentum space and the Fourier transform.

We can play the same game in the opposite direction of the Fourier transform, i.e., compute the ratio of the H4-extrapolated propagator in position space to the Fourier transform of the H4-extrapolated one in momentum space, $\Delta_{FT}(x)$. The results are plotted in Fig.~\ref{fig:ratioXP}(right), and show that there is a large bias for small distances in this case except for $\alpha_4=0.5$. In this case, we can also identify a non-negligible slope in $x^2$ for large values of $x^2$. This may correspond to a finite volume error in $\Delta_{H4}(x)$, although we cannot exclude a bias introduced by the inverse Fourier transform of ${\Delta}_{H4}(k)$.

Both figures provide the same message: only using the value $\alpha_4=1/2$ in the H4 extrapolation in position space, we obtain coherent results (from the point of view of the Fourier transforms) in position and momentum space.

\section{Application to Landau-gauge gluon propagator}
\label{sec:gluon}

\paragraph{Lattice set-up\\}

In order to show an application of the method proposed for a simple but not trivial case, we present in this section lattice calculations of gluon propagator in both position and momentum-space. One thousand SU(3) quenched gauge field configurations have been produced using Wilson gauge action for $\beta=6.0$ and $N=32$ lattice points in each space-time direction. Landau gauge has been fixed in the lattice through a Fourier accelerated method. Details on the lattice action and gauge fixing can be found in \cite{Ayala:2012pb,Boucaud:2018xup}. In order to write dimensional quantities in physical units, we will use the lattice-spacing provided by \cite{Necco:2001xg}, $a^{-1}\approx 2.045 {\rm GeV}$.

Once the gauge field configurations are in Landau gauge, the fields are defined from the link variables $U_\mu(x)$ as:
\begin{eqnarray}
A_\mu(x+\hat\mu/2) = \left.\frac{U_\mu(x)-U^\dagger_\mu(x)}{2 i a g_0}\right|_{traceless} 
\end{eqnarray}
and its momentum-space counterpart is obtained via a fast Fourier transform (FFT) as:
\begin{eqnarray}
{A}^a_\mu(k) = \frac{1}{2} {\rm Tr}\sum_x A_\mu(x+\hat\mu/2) \lambda^a 
e^{i k (x+\hat\mu/2)} \ .
\end{eqnarray}

The scalar component of momentum-space gluon propagator is obtained as:
\begin{eqnarray}
{\Delta}(k) =  \frac{1}{24} \langle \widetilde{A}^a_\mu(k)\widetilde{A}^a_\mu(-k) \rangle
\end{eqnarray}
where $\langle\cdots\rangle$ stands for average over gauge field configurations. We will obtain the diagonal contribution to the position-space propagator analogously as:
\begin{eqnarray}
\Delta(x) &=& \frac{1}{24 L^4} \sum_y \langle A^a_\mu(y) A^a_\mu(y-x)\rangle = \\
&=&  \frac{1}{24 L^4} \sum_y \sum_{k, k'} e^{-i k y } e^{-i k' (y-x) }\langle \widetilde{A}^a_\mu(k) \widetilde{A}^a_\mu(k')\rangle = \\
&=& \sum_k e^{-i k x } {\Delta}(k) 
\end{eqnarray}
which is the discrete inverse Fourier transform of the gluon propagator in momentum space.
Via this discrete Fourier transform we will obtain gluon propagator in position space for all the lattice sites which should be a rotationally invariant function of $x^2$ in the continuum limit.

\paragraph{Gluon propagator in momentum space\\}

\begin{figure}[h]
	\begin{center}
		\includegraphics{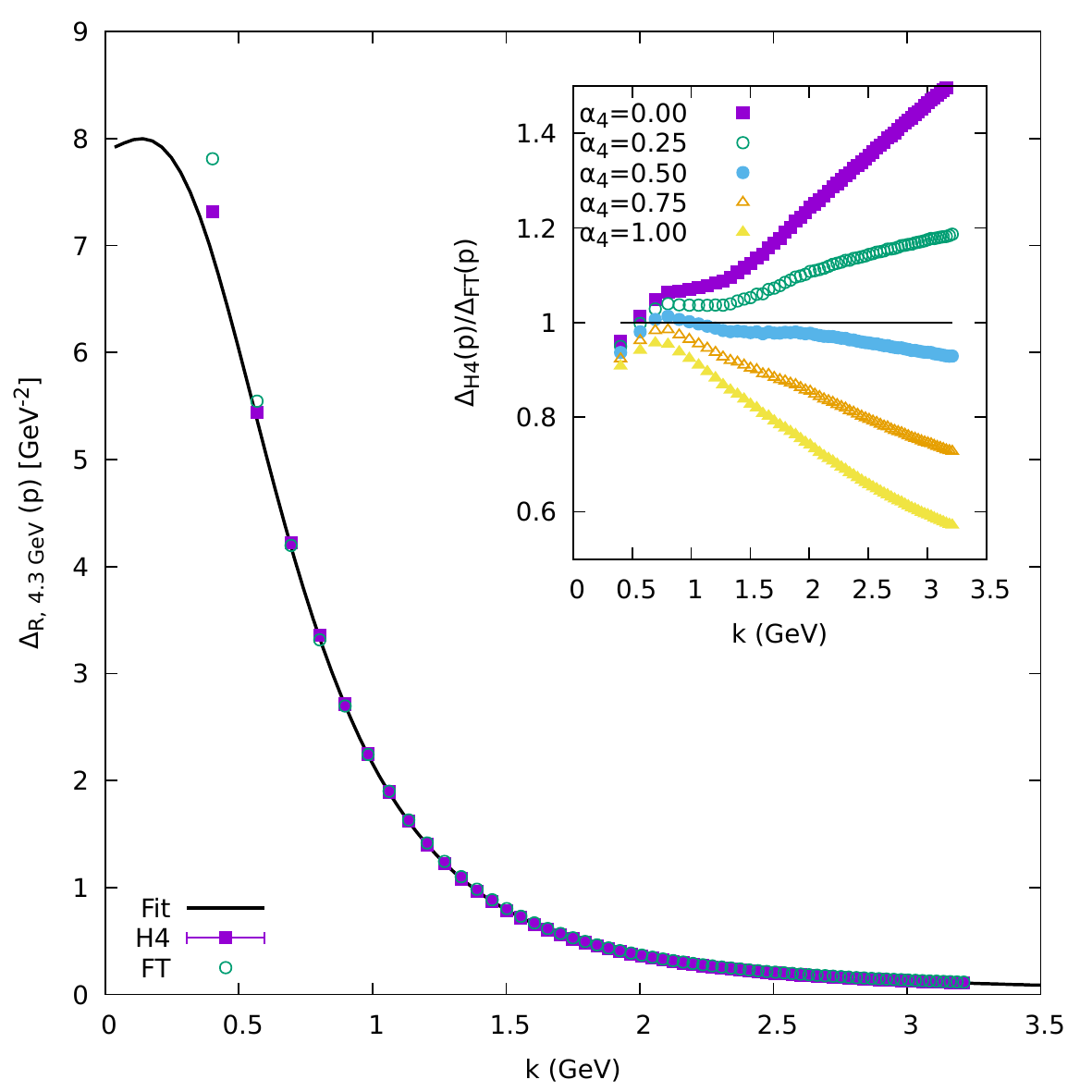}
	\end{center}\caption{Gluon propagator in momentum space renormalized at $\mu=4.3{\rm GeV}$ as a function of the momentum for the H4 extrapolated data in momentum space (purple full squares) and from the Fourier transform of the position-space ones (open green circles). For comparison, the parametrization proposed in \cite{Aguilar:2021okw} has been represented as the full line. The inset shows the ratio of the bare gluon propagators obtained from the application of H4 extrapolation in momentum space and the one coming from the Fourier transform of x-space ones. Each set corresponds to a different value of $\alpha_4$ for the extrapolations in position space (from top to bottom $\alpha_4=0$,  $\alpha_4=1/4$,  $\alpha_4=1/2$,  $\alpha_4=3/4$, and $\alpha_4=1$.)}
	\label{fig:gluonP}
\end{figure}

Gluon propagator in momentum-space after applying H4 extrapolation to reduce rotational symmetry breaking errors has been represented in Fig.~\ref{fig:gluonP}. For comparison, we have chosen $\mu=4.3$ GeV as the renormalization scale and added the precise parametrization discussed in Appendix B of \cite{Aguilar:2021okw}. There is a nice agreement except for the first point which is known to be affected by a finite volume error.

The inset shows the ratio of bare gluon propagators obtained from H4 extrapolation in momentum space and the one coming from the Fourier transform of the x-space one. Each set corresponds to a different value of the parameter $\alpha_4$ used in the position-space H4 extrapolation. As in the free-boson case, the case $\alpha_4=1/2$ seems to be preferred with a deviation from $1$ that is below $\sim 5\%$ for the whole accessible momentum window.

\paragraph{Gluon propagator in position space\\}

\begin{figure}[h]
	\begin{center}
		\includegraphics{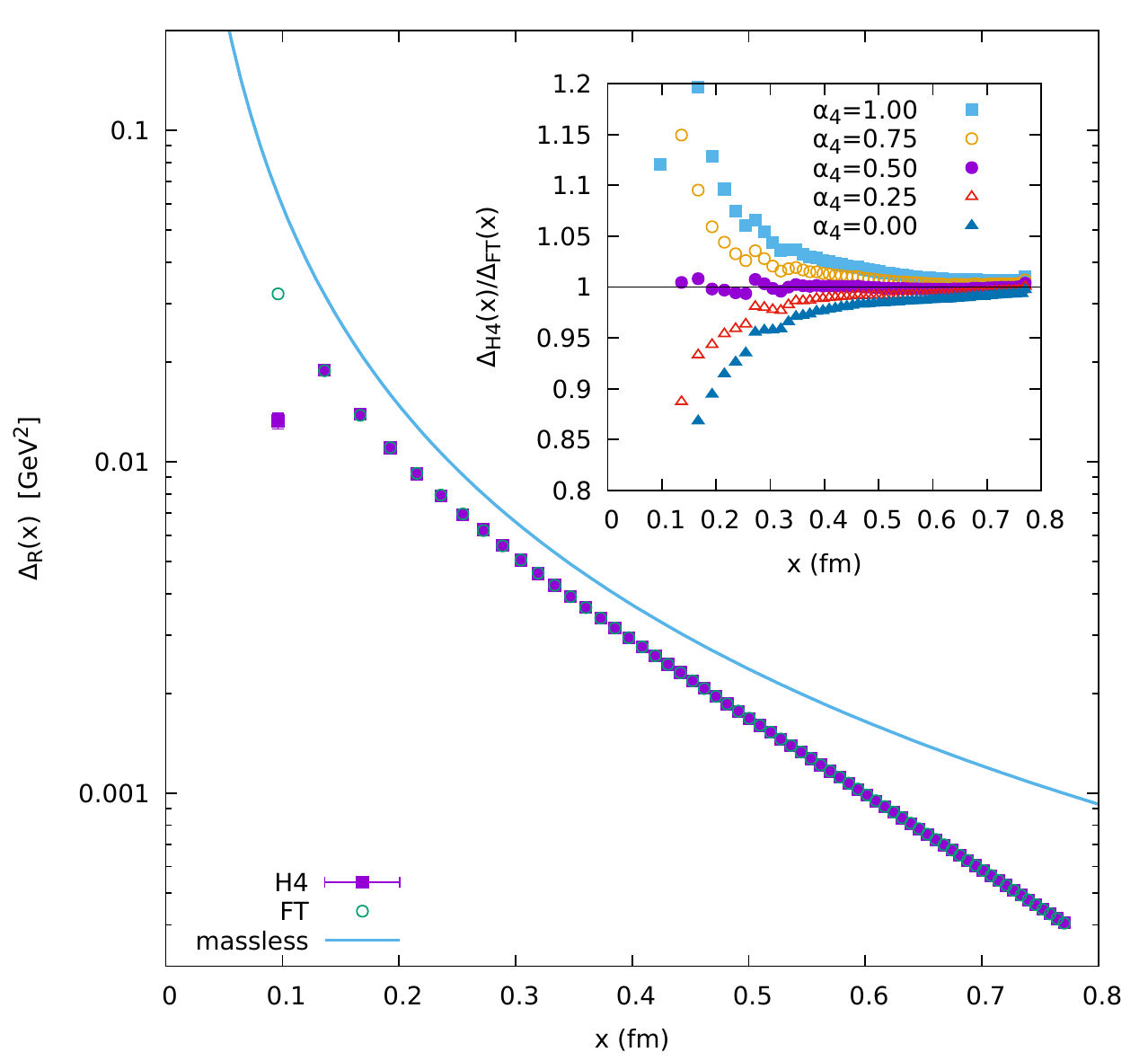}
	\end{center}\caption{Gluon propagator in position-space for the H4 extrapolated data in position (purple full squares) and from the Fourier transform of the momentum-space ones (open green circles, they are superimposed to the purple set except for the first point). The inset shows the ratio of the bare gluon propagators obtained from the application of H4 extrapolation in momentum space and the one coming from the Fourier transform of x-space ones. Each set corresponds to a different value of $\alpha_4$ for the extrapolations in position space (from top to bottom $\alpha_4=1$,  $3/4$,  $1/2$,  $1/4$, and $0$.).}
	\label{fig:gluonX}
\end{figure}

Our results in position space have been represented in Fig.~\ref{fig:gluonX}. 
For large distances ($x\gtrsim 0.5{\rm fm}$), the behavior of Landau gauge gluon propagator shows an exponential fall-off or Yukawa-like behavior \cite{Suganuma:2010mm} due to the dynamical gluon mass generation which is in marked contrast with the $1/x^2$ trend expected for a massless boson.
As in the previous case, the inset in Fig.~\ref{fig:gluonX} shows the ratio of the H4-extrapolated propagator in position space to the Fourier transform of the momentum-space one. We can appreciate in this case that the choice $\alpha_4=1/2$ is clearly preferred, with an overall deviation of this ratio from $1$ which is below $\sim 1\%$ for the whole set of positions. The only exception is the first point where the value provided by the H4 extrapolation method is far from the Fourier transform one.

\paragraph{Comparison with other methods\\}

\begin{figure}[h]
	\begin{center}
		\begin{tabular}{c c}
		\includegraphics[width=0.49\textwidth]{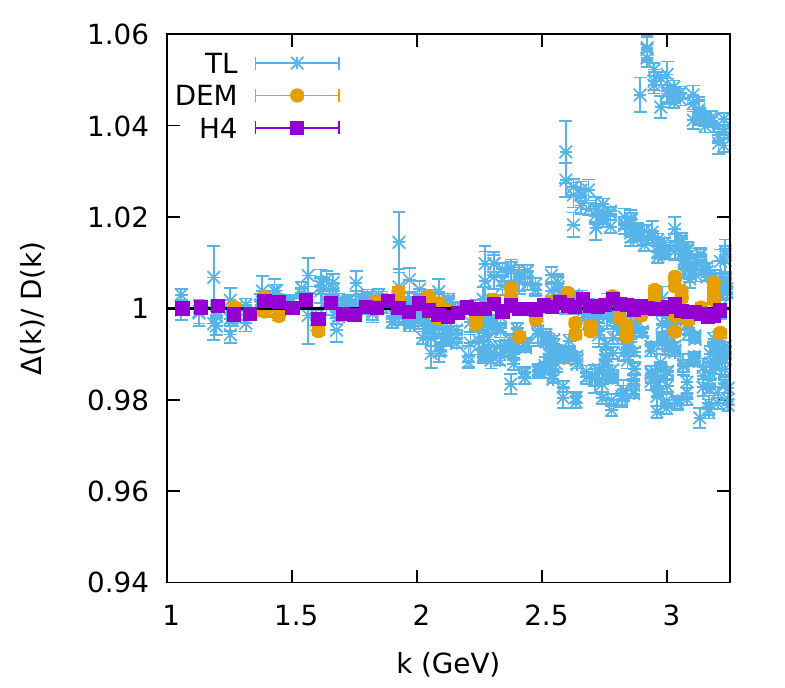} &
		\includegraphics[width=0.49\textwidth]{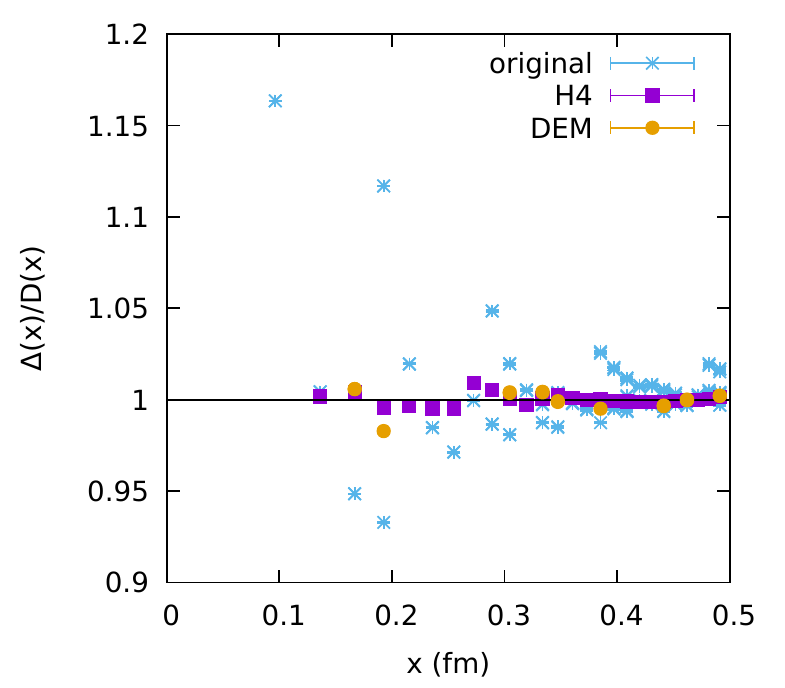}
		\end{tabular}
	\end{center}\caption{Ratio of the lattice data after treatment of the hypercubic artifacts and the continuum fitting functions in Eqs.~(\ref{eq:kfit},\ref{eq:xfit}) for gluon propagator in momentum-space (left) and position space (right).}
	\label{fig:methods}
\end{figure}

Let us finish with a comparison of the noise in the result of applying the H4 extrapolation, {\it democratic} method and the tree-level correction discussed in Section \ref{sec:kg}. For the purposes of making a direct comparison of the noise exhibited by the lattice data after application of these methods, we have fitted gluon propagator in momentum space to the functional form:
\begin{equation}\label{eq:kfit}
	{D}(k)=\frac{a k^2 + b}{k^4 + c k^2 + d}
\end{equation}
that appears, for example, in a Refined Gribov-Zwanziger framework~\cite{Dudal:2008sp,Dudal:2019vvt} and reproduces qualitatively very well the lattice data. In left panel of  Fig.~\ref{fig:methods} the ratio $\Delta(k)/D(k)$ has been plotted for each method (for each method the function $D(k)$ has been fitted separately and used for the ratio). The fluctuations around unity in this plot are due to the hypercubic errors that have not been fully eliminated by the method employed. It shows how in momentum space, H4-method clearly performs much better than the others, with noise at the level of $\sim 0.1\%$.

In position space, we have fitted the data for small distances to a function inspired by the asymptotic behavior of Eq.(\ref{KGcontinuumX}):
\begin{equation}\label{eq:xfit}
	{D}(x)=\frac{e^{-\xi x}}{x^{3/2}} \times (a + b x + c x^2)
\end{equation}
and represented the ratio $\Delta(x)/D(x)$ in the right panel of  Fig.~\ref{fig:methods}. In this case the noise is larger than in momentum space, but is kept below $2\%$ with the H4-method. The performance of the democratic method in this sense is similar except that the number of data points that survive the cylindrical cut is only a little fraction of the total number, while in the H4-method all the values of $n^2$ are used. Although the democratic method provides smooth data, they seem to be rather far from the continuum limit, at least it was the case for Klein-Gordon propagator (remind Fig.~\ref{fig:extra}).

It is important to remind that the conclusions of this paragraph are independent of the presence of O4-symmetric artifacts, that can only be eliminated using different lattice spacings and volumes or judiciously applying the Fourier transform method as sketched above.

\paragraph{Final comment\\}

There is one more comment in order with respect to the lattice errors (either discretization or finite volume). When analyzing the inset plots of Figs.~\ref{fig:gluonP} and \ref{fig:gluonX} for $\alpha_4=1/2$, one cannot distinguish without further analyses the errors coming from the H4-extrapolation results from those that have been introduced when computing the approximate Fourier transform described in section \ref{sec:ft}. From our experience in the free case, we can judge that the dependence in $k$ shown by the ratio in the inset plot of Fig.~\ref{fig:gluonP} must have been introduced by the Fourier transform from positions to momenta rather than by the H4-extrapolation in momentum space. This judgment, presented here without proof, is based in the fact that H4 extrapolation tends to work better in momentum than in position-space, for reasons already discussed along the manuscript, and in the small-x dependency of $\Delta(x)$, that diverges, what makes difficult to evaluate the inverse Fourier transform contribution from this region.

Finally, due to i) the better performance of the H4-extrapolation method in momentum space and ii) the difficulties of computing the approximate inverse Fourier transform of propagators that diverge at zero, the ratio presented in the inset of Fig.~\ref{fig:gluonX} seems better suited than the one in Fig.\ref{fig:gluonP} to analyze the possible presence of uncorrected O4 artifacts in the propagators.
Indeed, one may devise a more involved method for a relative callibration of the lattice spacing with a generalization of the one presented in \cite{Boucaud:2018xup} by combining both position and momentum-space propagators for different $\beta$'s and fix simultaneously the ratios of lattice spacings and the remaining $O4$ artifacts.

\section{Conclusions}

We have analyzed lattice errors appearing in boson propagators as a consequence of rotational invariance breaking. Contrarily to other lattice errors, the ones that are caused by breaking the rotational invariance can be efficiently removed using the spurious dependence of the propagators on higher order invariants of the discrete isometry group H(4). In order to accomplish this, we have proposed a generalization of the H4-extrapolation method presented in \cite{deSoto:2007ht} that can be straightforwardly applied both in momentum and position space. We have presented the method using the case of a free massive boson where the exact solution is known and then applied it to the Landau-gauge gluon propagator.

We have shown how the {\it democratic} orbits in position space are not the closest ones to the continuum results for a free-boson propagator, contrarily to the momentum-space case. This appears as the basic underlying reason for the failure of democratic methods in position space. The method proposed has the advantages that no {\it a priori} assumption is made over the $n^2$-dependence of the artifacts, and it does not assume any particular value of the ratio $n^{[4]}/(n^2)^2$ as being optimal. We have furthermore introduced the (approximate) continuum Fourier transforms as a method to check the remaining discretization errors after applying the H4 extrapolation. From a more phenomenological point of view, the combination of momentum and position-space lattice gluon propagators opens a new possible way of studying the deeply non-perturbative physics related to gluon mass generation in Yang-Mills theories.

\acknowledgments

The author is indebted to J. Rodr\'{\i}guez-Quintero for carefully reading this manuscript. The author acknowledges financial support from Spanish MICINN grant PID2019-107844-GB-C2, and regional Andalusian project P18-FR-5057.
The author acknowledges, too, the use of the computer facilities of C3UPO at Pablo de Olavide University.

\providecommand{\href}[2]{#2}\begingroup\raggedright\endgroup
\bibliography{refs}

\bibliographystyle{JHEP} 

\end{document}